\documentclass[twocolumn]{aastex62}

\received{2020 February 21}
\revised{2020 May 29}
\accepted{2020 June 1}

\shorttitle{A SN candidate in \xmm\ data}
\shortauthors{G. Novara et al.}

\begin{document}

\title{A supernova candidate at $z=0.092$ in \xmm\ archival data}


\correspondingauthor{Giovanni Novara}
\email{giovanni.novara@iusspavia.it}
\author[0000-0001-7304-9858]{Giovanni Novara}
\affil{Scuola Universitaria Superiore IUSS Pavia, Palazzo del Broletto, piazza della Vittoria 15, 27100 Pavia, Italy}
\affil{INAF--Istituto di Astrofisica Spaziale e Fisica Cosmica di Milano, via A.\,Corti 12, 20133 Milano, Italy}
\correspondingauthor{Paolo Esposito}
\author[0000-0003-4849-5092]{Paolo Esposito}
\email{paolo.esposito@iusspavia.it}
\affil{Scuola Universitaria Superiore IUSS Pavia, Palazzo del Broletto, piazza della Vittoria 15, 27100 Pavia, Italy}
\affil{INAF--Istituto di Astrofisica Spaziale e Fisica Cosmica di Milano, via A.\,Corti 12, 20133 Milano, Italy}
\author[0000-0002-6038-1090]{Andrea Tiengo}
\affil{Scuola Universitaria Superiore IUSS Pavia, Palazzo del Broletto, piazza della Vittoria 15, 27100 Pavia, Italy}
\affil{INAF--Istituto di Astrofisica Spaziale e Fisica Cosmica di Milano, via A.\,Corti 12, 20133 Milano, Italy}
\affil{Istituto Nazionale di Fisica Nucleare (INFN), Sezione di Pavia, via A.\,Bassi 6, 27100 Pavia, Italy}
\author[0000-0002-2553-0839]{Giacomo Vianello}
\affil{Department of Physics, Stanford University, 382 via Pueblo Mall, Stanford, CA 94305-4013 USA}
\author[0000-0002-9393-8078]{Ruben Salvaterra}
\affil{INAF--Istituto di Astrofisica Spaziale e Fisica Cosmica di Milano, via A.\,Corti 12, 20133 Milano, Italy}
\author[0000-0002-2526-1309]{Andrea Belfiore}
\affil{INAF--Istituto di Astrofisica Spaziale e Fisica Cosmica di Milano, via A.\,Corti 12, 20133 Milano, Italy}
\author[0000-0001-6739-687X]{Andrea De Luca}
\affil{INAF--Istituto di Astrofisica Spaziale e Fisica Cosmica di Milano, via A.\,Corti 12, 20133 Milano, Italy}
\affil{Istituto Nazionale di Fisica Nucleare (INFN), Sezione di Pavia, via A.\,Bassi 6, 27100 Pavia, Italy}
\author[0000-0001-7164-1508]{Paolo D'Avanzo}
\affil{INAF--Osservatorio Astronomico di Brera, via E.\,Bianchi 46, 23807 Merate (LC), Italy}
\author[0000-0002-9875-426X]{Jochen Greiner}
\affil{Max-Planck-Institut f\"ur extraterrestrische Physik (MPE), Giessenbachstrasse 1, 85740 Garching, Germany}
\author[0000-0002-2282-5850]{Marco Scodeggio}
\affil{INAF--Istituto di Astrofisica Spaziale e Fisica Cosmica di Milano, via A.\,Corti 12, 20133 Milano, Italy}
\author{Simon Rosen}
\affil{European Space Astronomy Center (ESA/ESAC), Operations Department, Vilanueva de la Ca\~nada, 28692 Madrid, Spain}
\affil{Department of Physics and Astronomy, University of Leicester, LE1 7RH Leicester, UK}
\author{Corentin Delvaux}
\affil{Max-Planck-Institut f\"ur extraterrestrische Physik (MPE), Giessenbachstrasse 1, 85740 Garching, Germany}
\author[0000-0001-8646-4858]{Elena Pian}
\affil{INAF--Osservatorio di Astrofisica e Scienza dello Spazio, via P.\,Gobetti 101, 40129 Bologna, Italy}
\author[0000-0001-6278-1576]{Sergio Campana}
\affil{INAF--Osservatorio Astronomico di Brera, via E.\,Bianchi 46, 23807 Merate (LC), Italy}
\author{Gianni Lisini}
\affil{Scuola Universitaria Superiore IUSS Pavia, Palazzo del Broletto, piazza della Vittoria 15, 27100 Pavia, Italy}
\author[0000-0003-3259-7801]{Sandro Mereghetti}
\affil{INAF--Istituto di Astrofisica Spaziale e Fisica Cosmica di Milano, via A.\,Corti 12, 20133 Milano, Italy}
\author[0000-0001-5480-6438]{G. L. Israel}
\affil{INAF--Osservatorio Astronomico di Roma, via Frascati 33, 00078 Monteporzio Catone, Italy}
\def\xmm {\emph{XMM--Newton}}
\def\cxo {\emph{Chandra}}
\def\swift {\emph{Swift}}
\def\src {\mbox{EXMM\,023135.0--603743}}
\def\flux {\mbox{erg cm$^{-2}$ s$^{-1}$}}
\def\lum {\mbox{erg s$^{-1}$}}
\def\nh {$N_{\rm H}$}

\begin{abstract}
During a search for X-ray transients in the \xmm\ archive within the EXTraS project, we discovered a new X-ray source that is detected only during a $\sim$5\,min interval of a $\sim$21\,h-long observation performed on 2011 June 21 (\object{EXMM\,023135.0--603743},  probability of a random Poissonian fluctuation: $\sim$$1.4\times10^{-27}$). With dedicated follow-up observations, we found that its position is consistent with a star-forming galaxy ($\rm SFR = 1$--$2\,M_\odot$\,yr$^{-1}$) at redshift $z=0.092\pm0.003$ ($d=435\pm15$\,Mpc). At this redshift, the energy released during the transient event was $2.8\times10^{46}$\,erg in the 0.3--10 keV energy band (in the source rest frame). The luminosity of the transient, together with its spectral and timing properties, make \src\ a gripping analog to the X-ray transient associated to SN\,2008D, which was discovered during a \emph{Swift}/XRT observation of the nearby ($d=27$\,Mpc) supernova-rich galaxy NGC\,2770. We interpret the \xmm\ event as a supernova shock break-out or an early cocoon, and show that our serendipitous discovery is broadly compatible with the rate of core-collapse supernovae derived from optical observations and much higher than that of tidal disruption events.\\
\end{abstract}

\section{Introduction} \label{sec:intro}

High energy transients for the most part are discovered in the hard-X/gamma-ray band by instruments monitoring a large fraction of the sky. In the soft X-ray band ($E<10$\,keV), instead, the most sensitive instruments have small fields of view (less than a few tenths of a squared degree). However, some missions carrying narrow-field X-ray instruments have spent a long time in orbit, accumulating many years of exposure time, which makes it possible the serendipitous discovery of rare transient events during observations of unrelated targets. In particular, \xmm\ has been in orbit since December 1999 and has the largest effective area among current imaging X-ray telescopes. It is therefore the ideal mission to search for faint transients in the soft X-ray band. 

One of the objectives of EXTraS,\footnote{Exploring the X-ray Transient and variable Sky; see \url{http://www.extras-fp7.eu}.} an European-Union-funded project aimed at mining the \xmm\ archival data in the time domain \citep{deluca16}, 
was 
the identification of 
short-lived transient X-ray sources. In particular, we developed an algorithm to search for new point sources that were sufficiently bright only for a small fraction of the observation and could not be detected by a standard analysis of the full exposure. Such X-ray sources are therefore not included in the \xmm\ serendipitous source catalogues released by the \xmm\ Survey Science Centre\footnote{See \url{http://xmmssc.irap.omp.eu/}.} \citep{rosen16}.

We performed a systematic analysis of all the observations used for the 3XMM-DR5 catalogue \citep{rosen16} and, after the careful screening of the results, we derived a catalogue of 136 new transients with a duration $<$5\,000\,s, which is publicly available through the EXTraS online archive.\footnote{See the EXTraS Transient Catalogue at \url{http://www88.lamp.le.ac.uk/extras/query/extras\_transients}.} Among these transients, \src\ is the one with the shortest duration (315\,s) and is the subject of this work.
In Section\,2, we describe the procedures that allowed us to discover this transient. In Section\,3, we
report on the results of the timing and spectral analysis of the X-ray data, while Section\,4 is devoted to the follow-up optical observations we carried out. 
The discussion of the nature of the transient follows in Section\,5, together with some considerations on the rate of such events and the perspectives for future missions.


\section{EXTraS pipeline to search for X-ray transients} 
\subsection{Data preparation}
The European Photon Imaging Camera (EPIC) on board \xmm\ consists of one pn \citep{struder01} and two MOS \citep{turner01} CCD cameras sensitive to photons with energy between 0.2 and 12\,keV. Each camera is installed behind an X-ray telescope with 58 nested grazing-incidence mirrors and focal length of 7.5\,m.

The EPIC data were processed with version 14.0.0 of the Scientific Analysis Software (SAS).
The analysis was performed only on events with valid pattern (0--4 for the pn and 0--12 for the MOS) and FLAG=0 (to avoid pixels close to CCD boundaries and dead columns). In contrast to standard analysis, the search for new X-ray sources was performed without the exclusion of the time periods in which the particle background was particularly high.  

\subsection{The EXTraS procedure}\label{procedure}
The EXTraS process aimed at the discovery of new transients consists of the division of each EPIC observation into sub-exposures and in the search for
new point sources that might have been bright only for short time intervals. In order to search over a broad range of time scales, the time-resolved source detection is 
applied to time intervals of 
variable duration, 
determined through a preliminary search for an excess of counts in limited time periods in small regions of the detector. 

This step of the analysis is performed
using the Bayesian Blocks (BB) algorithm \citep{scargle98,scargle13} in the 0.2--12\,keV; 0.5--2\,keV; 2--10\,keV energy bands. This adaptive-binning algorithm finds statistically-significant changes in the count rate by maximizing the fitness function for a piecewise-constant representation of the data, starting from an event list. 

To reduce the number of spurious detections and to sample a broad range of time intervals, we modified the BB algorithm to account for changes in the background rate. The new algorithm can deal with highly-variable background such as that found in \xmm\ data during soft-proton flares. For each observation,  the field of view is divided in partially-overlapping $30''\times30''$ box regions and the BB algorithm is run on each of them. 
Regions with no significant variability with respect to the local background light curve return only one \emph{block} (a time bin) covering the whole observation, while regions containing candidate transients return more blocks.

To evaluate properly the background light curve and minimize the contribution from the possible variability of known sources, the BB algorithm excludes regions of source-intensity-dependent size around the
point sources detected in the full observation.
To examine also these regions, where interesting transients might appear (especially in crowded X-ray fields, such as star-forming regions and nearby galaxies), we developed a specific algorithm. For each observation, it creates images integrated over a fixed time interval (1000\,s) of regions with side of $40''$ around the sources excluded by the BB algorithm and tests for the presence of excesses in addition to known sources on a grid of fixed positions by a sliding-cell search. After performing this analysis on each time bin, all intervals where the same source was active are merged.
Among the time segments identified either in this way or by the BB analysis, we selected only those with duration shorter than 5\,ks (the minimum duration of standard EPIC exposures) and coming from regions with a spatial distribution of the events that is better fit (at $>$5\,$\sigma$ confidence level) with the addition of a point source rather than by a simple isotropic background. 

In the time interval obtained from the merged segments, a source detection based on the SAS task emldetect is performed on the combined EPIC MOS and pn images  
accumulated in the five standard 3XMM energy bands (0.2--0.5, \mbox{0.5--1}, 1--2, 2--4.5, and 4.5--12\,keV).
The sources detected with $\rm DET\_ML>6$ in the cumulative \mbox{0.2--12\,keV}
band\footnote{This is the standard detection threshold adopted for the \xmm\ source catalogues; $\mathrm{DET\_ML} = - \ln(p)$, where $p$ is the probability that the count excess was due to a random Poissonian fluctuation \citep[see][and \url{http://xmm-tools.cosmos.esa.int/external/sas/current/doc/emldetect/node3.html} for an extensive description of the parameter]{cash79}.\label{detmlnote}} 
are compared with the reference source list  for the whole observation, 
looking for new point-like X-ray sources.

\begin{figure*}
\centering
\resizebox{\hsize}{!}{\includegraphics[angle=0]{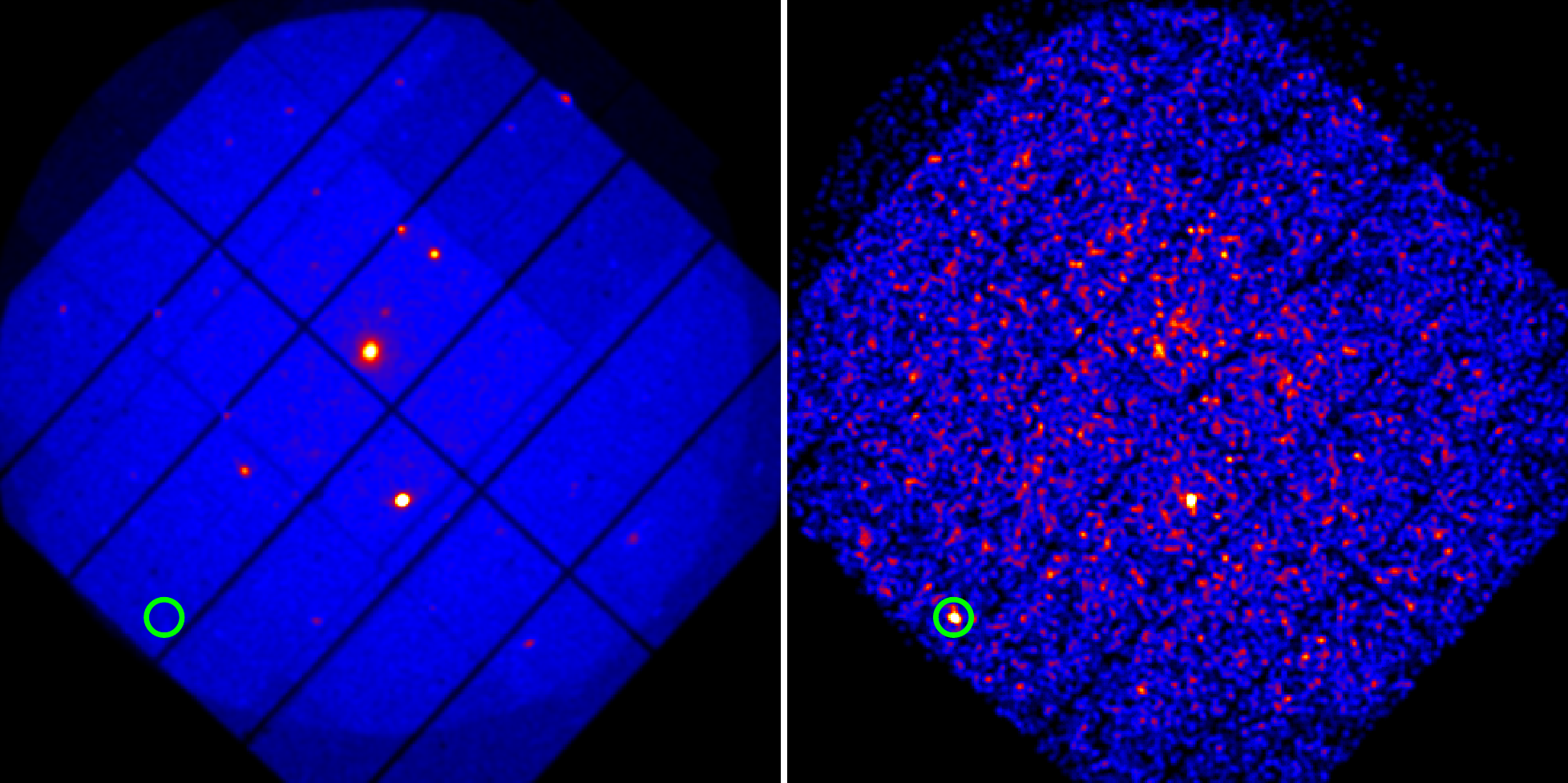}}
\caption{\label{figXMM} EPIC images (0.2--12\,keV, combined pn and MOS data) with Gaussian smoothing accumulated from the full exposure (left panel) and only during the 315\,s time interval in which the transient was detected (right panel). 
The green circles (40$''$ radius) indicate the position of the X-ray transient.}
\end{figure*}

After the exclusion of bright pixels and particle tracks,
we obtained about one thousand transient candidates from $\sim$7800\, 3XMM-DR5 observations.
The candidates with the largest likelihood of detection $\rm DET\_ML$
(we set the threshold at $\rm DET\_ML > 15$, leading to 596 sources) were visually screened to exclude spurious detections or persistent sources erroneously classified as transients. 
The result of this screening process is the publicly available EXTraS transient catalogue, which contains 136 new transient X-ray sources.
Among them, \src\ was detected in the shortest time interval (315\,s; see Fig.\,\ref{figXMM}). 

\section{\src: \xmm\ data analysis and results}
\subsection{Detection parameters and astrometry}
In the observation where \src\ was discovered (obs.ID:\,0675010401, started on 2011 June 21, exposure time of approximatively 77\,ks, see Table\,\ref{tab:SNparameters}), the two MOS and the pn cameras were set in Full Frame mode (time resolution of 2.6\,s and 73.4\,ms, respectively); all detectors were operated with the thin optical-blocking filter. 

The main  parameters characterizing the detection of \src\ are reported in Table\,\ref{tab:SNparameters}.
To refine its position, we performed an astrometric correction by cross-matching the brightest sources detected in the 
\xmm\ observation with the USNO\,B1 optical catalogue \citep{monet03}. 
The IRAF tasks geomap and geoxytran were used to estimate the roto-translation between the X-ray and USNO B1 optical coordinates of the reference sources and calculate the corrected X-ray coordinates of \src: $\rm RA=02^h31^m34\fs9$, $\rm Dec=-60^{\circ}37'43\farcs3$ (J2000). The 1$\sigma$ error radius, which was calculated as the sum in quadrature of the systematic astrometric error ($1\farcs3$ rms) and the statistical error on the coordinates measured for the source, is $1\farcs9$.
Within this uncertainty, the position of \src\ is consistent with a blue galaxy visible in the 2nd Digitized Sky Survey images and with no redshift reported in literature.
\begin{deluxetable}{ccccc}[h!]
\tablecaption{Parameters of \src\ in observation 0675010401. The detection likelihood is indicated by DET\_ML (see note \ref{detmlnote}), while SCTS is the number of net counts. 
\label{tab:SNparameters}}
\tablecolumns{5}
\tablenum{1}
\tablewidth{0pt}
\tablehead{
\colhead{Parameter} &
\colhead{EPIC} &
\colhead{pn} &
\colhead{MOS1} & 
\colhead{MOS2}
}
\startdata
DET\_ML\tablenotemark{a} & 61.8 & 30.3 & 13.8 & 20.1 \\
SCTS\tablenotemark{a} & 54.7 & 27.3 & 11.1 & 16.3 \\
Exposure time (ks) & -- & 60.2 & 75.8 & 76.9 \\
\enddata
\tablenotetext{a}{In the 0.2--12\,keV energy range and only in the 315\,s time interval in which \src\ was detected.}
\end{deluxetable}
%
\subsection{Light curve and spectrum}
\begin{figure*}
\centering
\resizebox{\hsize}{!}{\includegraphics[angle=0]{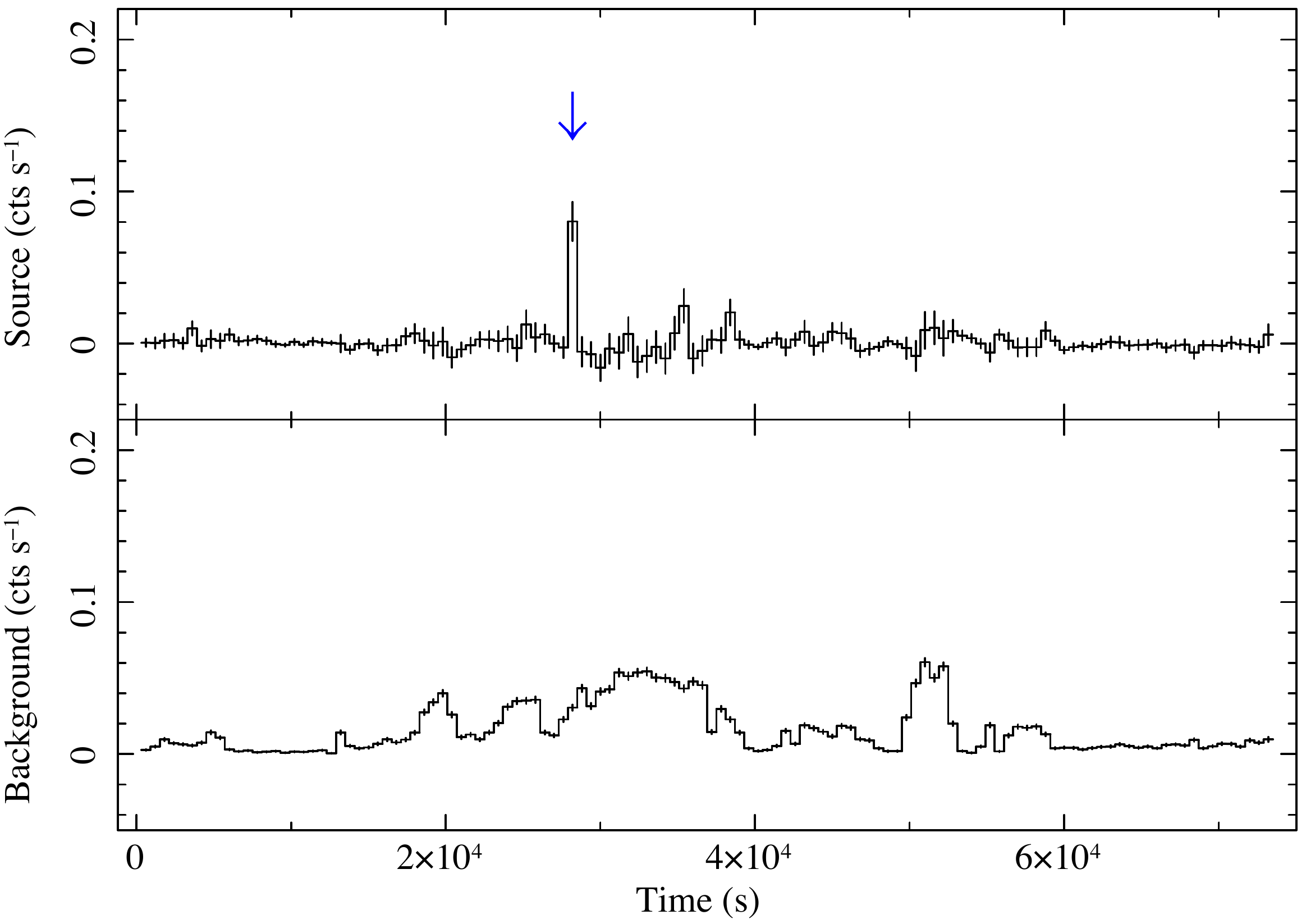}\includegraphics[angle=0]{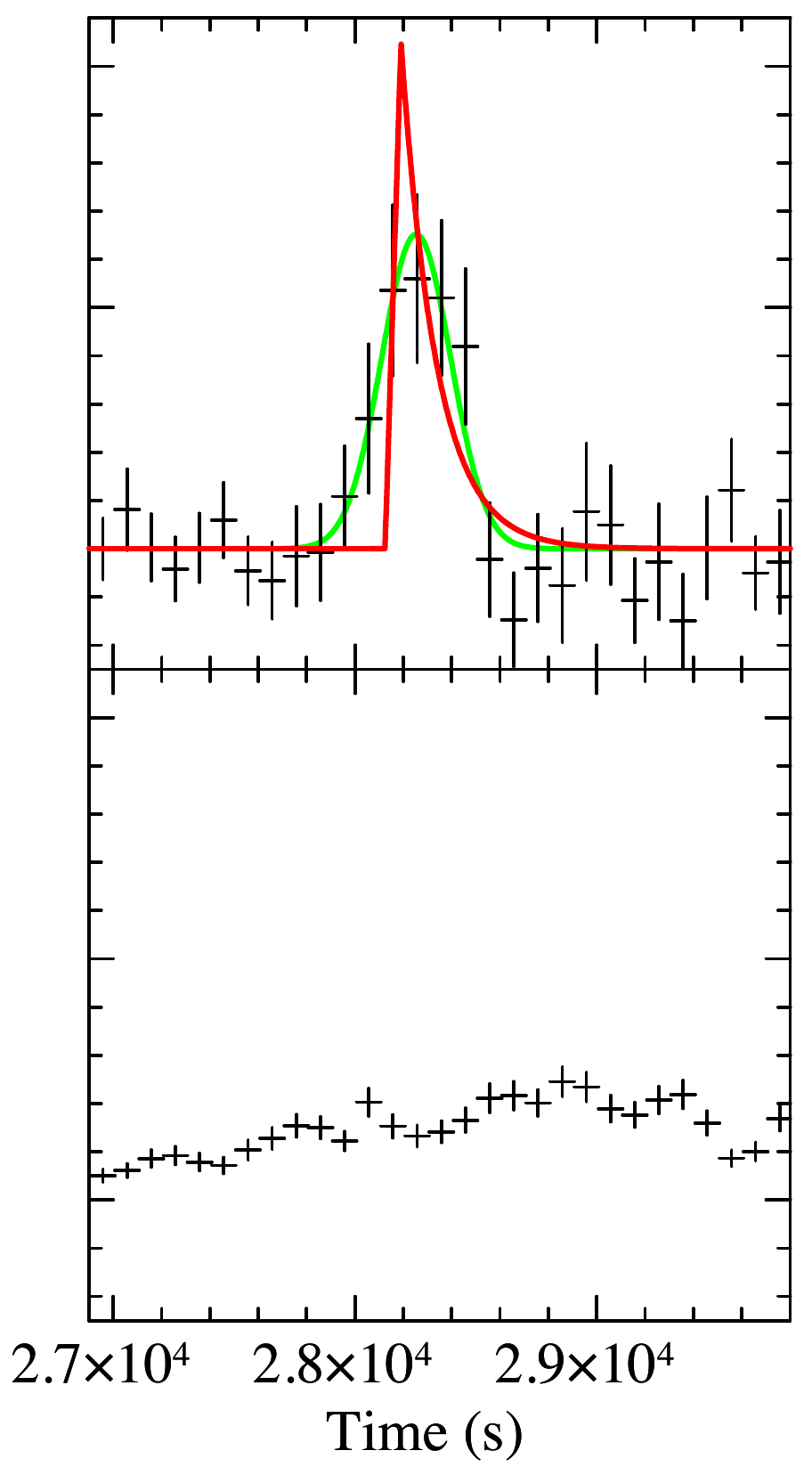}}
\caption{\label{figLC} {\it Left panel, top:} EPIC background-subtracted light curve extracted from a 20\,arcsec radius circle in the 0.5--5\,keV energy band with 10 minutes time bins. The blue arrow indicates the flare of \src. {\it Left panel, bottom:} corresponding background light curve extracted from a nearby circular region with 60\,arcsec radius and rescaled to the area of the source extraction region. {\it Right panel:} Same as on the left, but with 100-s  time bins and closing up on the source. The lines show the fits with a Gaussian (green) and a FRED (red) model (see Sect.\,\ref{sec:5.1} for details); in the latter, the burst rise time and e-folding decay time are fixed to the best-fit values derived for SN\,2008D in \citet{soderberg08}.}
\end{figure*}

The EPIC background-subtracted light curve of the source and the background (extracted from circles with radii of 20 and 60 arcsec, respectively) in the 0.5--5 keV energy band (the band in which the signal-to-noise ratio is maximized) are shown in Fig.\,\ref{figLC}. Although the background is very strong and variable, a significant count-rate excess is visible $\sim$28\,ks after the start of the observation. As shown in Fig.\,\ref{figLC}, this flare can be fit with a Gaussian centered at 2011 June 21 18:50:$54\pm30$\,s UT and with $\sigma=(146\pm22)$\,s. Integrating the Gaussian, we derive that the flare is formed by 47\,cts in the \mbox{0.5--5\,keV} band. The addition of a constant component to the model is not required, with a 3$\sigma$ upper limit of \mbox{$7.4\times10^{-4}$\,cts\,s$^{-1}$}, which means that the emission outside the flare is perfectly consistent with the background level. More complicate, asymmetric models are not required to describe the flare (see also the discussion in Sect.\,\ref{sec:5.1}).

The X-ray spectrum of the transient was extracted from a circle with 40 arcsec radius 
in the 315\,s interval during which it was detected and the corresponding background spectrum from a nearby source-free region.
The spectral analysis was carried out with the fitting package XSPEC and adopting the abundances by \citet{wilms00}.
Using a power-law model absorbed both in our Galaxy (we fixed the column value in the XSPEC model component tbabs to $N_{\rm H,Gal}=3\times10^{20}$\,cm$^{-2}$; \citealt{dickey90}, \citealt{kalberla05}) and in the host galaxy (ztbabs with $z=0.092$, see Sect.\,\ref{ctioblanco}), we derive a photon index $\Gamma=2.6^{+0.7}_{-0.6}$ and a $>$95\% evidence for higher-than-Galactic absorption, indicating a local component in the host galaxy, $N_{\rm H,z}=(1.0^{+0.6}_{-0.5})\times10^{22}$\,cm$^{-2}$. The goodness of the fit,  evaluated as the percentage of Monte Carlo realizations that had Cash statistics \mbox{(C-stat)} lower than the best-fitting C-stat, is 86\% (we performed $10^4$ simulations).
The average luminosity (for $d=435$\,Mpc, see Sect.\,\ref{ctioblanco}) was $\approx$$9\times10^{43}$\,\lum\ in the 0.3--10\,keV band (in the source rest frame). Therefore, the total energy of the flare was  $2.8\times10^{46}$\,erg. The peak luminosity was \mbox{$1.2\times10^{44}$\,\lum} and the 3$\sigma$ upper limit on the persistent X-ray luminosity outside the flare interval is $2.8\times10^{41}$\,\lum. 
A somewhat better fit (76\% of the realizations with C-stat lower than the best-fit one) is obtained with a blackbody model. The temperature is equivalent to $0.57^{+0.10}_{-0.08}$\,keV and for the local absorption we obtain $N_{\rm H,z}=1.5\times10^{21}$\,cm$^{-2}$, but the value is loosely constrained and consistent with zero. Assuming this model, the total energy emitted was $8\times10^{45}$\,erg.

\section{Follow-up optical observations of \src}\label{followups}
\subsection{CTIO/Blanco telescope observations}\label{ctioblanco}
\begin{figure}
\resizebox{\hsize}{!}{\includegraphics[angle=0]{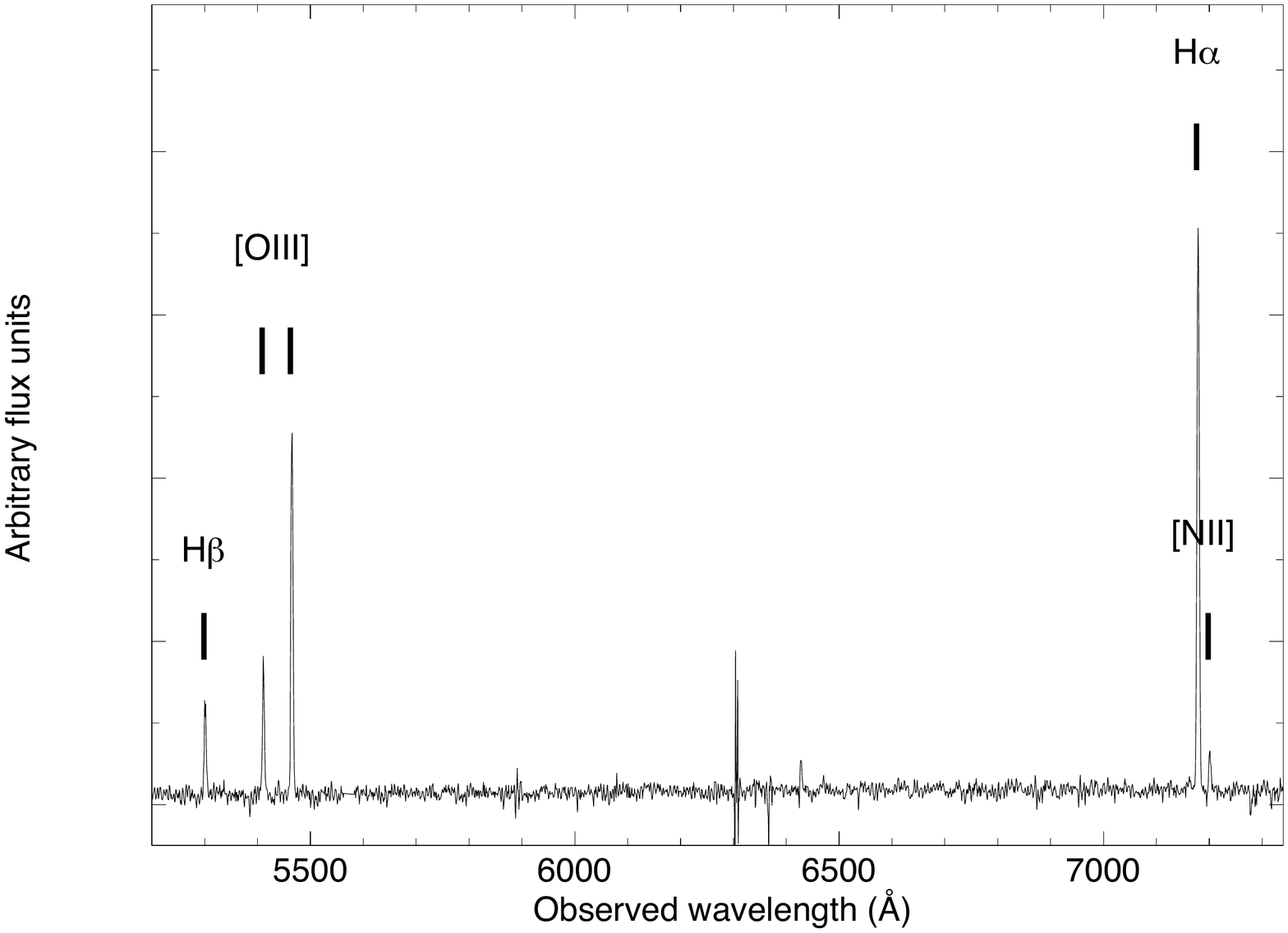}}
\caption{\label{figctio_spe}CTIO/Blanco telescope optical spectrum of the host galaxy of EXMM023135.0–603743 at $z = 0.092 \pm 0.002$.}
\end{figure}

We observed the field of \src\ with the 4-m CTIO/Blanco telescope equipped with the COSMOS spectrograph. A series of three consecutive spectra, each one lasting 900\,s, was obtained on 2016 July 12 starting at 09:20 UT. The spectroscopic observations were executed using a 1.3 arcsec slit, the r2K grism (GG455 filter), covering the 4955--9023\,\AA\ wavelength range with a dispersion of 1 \AA/pixel. The spectral reduction and extraction were carried out using standard procedures under the ESO-MIDAS\footnote{See \url{http://www.eso.org/projects/esomidas/}.} package. The wavelength calibration has been checked against sky emission lines. From the detection of several emission (H$\beta$, [O\,\textsc{iii}], H$\alpha$, [N\,\textsc{ii}]) and
absorption lines, we derived a redshift $z = 0.092 \pm 0.002$ (Fig.
\ref{figctio_spe}).  With the cosmological parameters in \citet[][assumed throughout the paper]{planck16}, this redshift corresponds to a luminosity distance $d=435\pm15$\,Mpc \citep{wright06}.
%
%
%

\subsection{La Silla (MPG/ESO 2.2m GROND)}
\begin{figure}
\resizebox{\hsize}{!}{\includegraphics[angle=0]{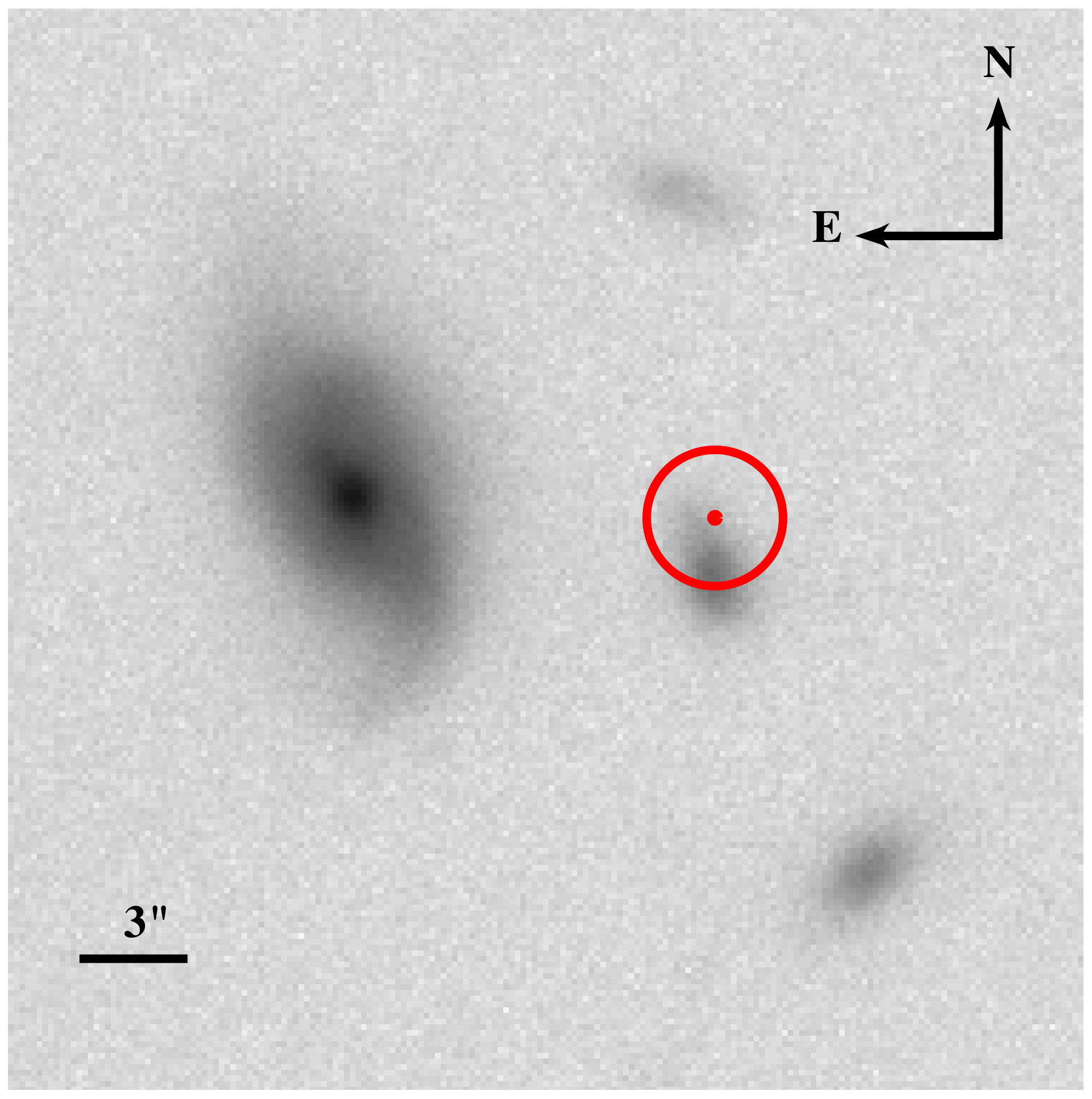}}
\caption{\label{figGROND}GROND optical image (30$''$ side). The red circle indicates the position of the X-ray transient and its  1$\sigma$ uncertainty (1.9$''$ radius) derived after astrometric correction against the USNO catalogue (the position of 5 common X-ray and optical sources have been used to register the image). Note that also the astrometry of the GROND image was performed using the USNO catalogue. The only object for which we could find an entry in a catalogue is the galaxy east (at $\sim$12\arcsec) of the transient, PGC361816 in the HyperLEDA data base \citep{makarov14}.}
\end{figure}

The transient field was observed with the seven-channel imager GROND \citep{greiner08} on 2016 July 30 starting at 08:30 UT at $\sim$1\arcsec\ seeing (Fig.\,\ref{figGROND}). The effective exposure was 36 minutes in $g'r'i'z'$, and 30 minutes in $JHK$. The GROND data were reduced in the standard manner \citep{kruhler08} using pyraf/IRAF \citep{tody93,yoldas08}. The optical/NIR imaging was calibrated against the \emph{Gaia} SkyMapper Southern Sky Survey\footnote{See \url{http://skymapper.anu.edu.au}.} \citep{gaia18,wbo18} catalogs for $g^\prime r^\prime i^\prime z^\prime$, and the 2MASS catalog \citep{skrutskie06} for the $JHK$ bands. This results in typical absolute accuracies of $\pm$0.03~mag in $g^\prime r^\prime i^\prime 
z^\prime$ and $\pm$0.05~mag in $JHK$.
\begin{deluxetable}{lc}
\tablecaption{GROND photometry. All magnitudes are in the AB system and are corrected for the foreground Galactic extinction. 
\label{tab:grond}}
\tablecolumns{2}
\tablenum{2}
\tablewidth{0pt}
\tablehead{
\colhead{Photometric band} &
\colhead{Magnitude} 
}
\startdata
$g'$ & $19.58\pm0.01$ \\
$r'$ & $19.58\pm0.01$ \\
$i'$ & $19.53\pm0.01$ \\
$z'$ & $19.67\pm0.02$\\
$J$  & $19.39\pm0.07$ \\
$H$  & $19.63\pm0.14$ \\
$K$  & $19.67\pm0.18$
\enddata
\end{deluxetable}
From the available optical and near-infrared photometry (Table\,\ref{tab:grond}), and the optical spectrum, it is possible to derive some basic physical properties of the host galaxy. We have used the GOSSIP spectral energy distribution (SED) fitting package \citep{franzetti08} to model the SED of the galaxy, starting from the multiband photometric information, and using the PEGASE2 population synthesis  models \citep{fioc97} coupled with a family of delayed star formation histories \citep{gavazzi02} to derive the spectral templates.
From the properties of the best fitting SED model, we estimate a galaxy stellar mass of $\approx$(2--$3)\times10^8$\,$M_{\odot}$, an absolute $B$-band magnitude of $-$18.49, and a star formation rate $\mathrm{SFR} \approx 1\,M_{\odot}$\,yr$^{-1}$. 
For and independent check, we also used the Le\,PHARE \citep{arnouts99,ilbert06} package with 70 templates from both the PEGASE2 and BC03 \citep{bruzual03} and we found that predictions of the best-fitting individual templates fall within the ranges (1.5--$3)\times10^8$\,$M_{\odot}$ for the mass and 0.1--$1\,M_{\odot}$\,yr$^{-1}$ for the SFR.

A direct measurement of the galaxy SFR can be obtained also from the H$\alpha$ line luminosity, as measured from the galaxy optical spectrum. To have a reliable absolute flux calibration for the spectrum, we scaled the spectroscopic fluxes so that the magnitude one can derive by integrating the spectroscopic flux over the $r$-band photometric passband matches the $r$-band magnitude from the GROND photometry. After this rescaling, the observed flux of the H$\alpha$ line, as derived by fitting a combination of a Gaussian profile for the emission line and a second order polynomial for the underlying background, is of 
$9.8\times10^{-15}$\,\flux, corresponding to a luminosity of $2.2\times10^{41}$\,\lum.
From this luminosity, following the recipe by \citet{kennicutt98}, we can derive an estimate of the current star formation rate of $\mathrm{SFR} =1.7\,M_{\odot}$\,yr$^{-1}$.

\section{Discussion}
\subsection{An X-ray flare associated to a supernova}\label{sec:5.1}
The energy released in the flare, its duration and spectrum, as well as the properties of the host galaxy, recall the characteristics of the X-ray transient associated to SN\,2008D in NGC\,2770, which was observed with {\it Neil Gehrels Swift Observatory}'s XRT and interpreted as the X-ray emission from the shock break-out of a core-collapse supernova \citep[][see also \citealt{modjaz09}]{soderberg08}. To explore further this analogy, we fit to the light curve of the transient a fast-rise-and-exponential-decay (FRED) model with the rise time ($t_{\rm r}=72$\,s) and $e$-folding decay time ($t_{\rm d}=129$\,s) fixed to the values reported in \citet{soderberg08}, obtaining a 
marginally acceptable fit 
(see Fig.\,\ref{figLC}). Even though the shape of the light curve is not necessarily a fingerprint of the event, to test the compatibility of the time evolution of these two supernova shock break-out candidates, we applied the two sample Kolmogorov--Smirnov (KS) test to the arrival times of their X-ray events, for a time interval of 600\,s starting from the beginning of the flares. Selecting the EPIC events in the energy band \mbox{0.5--5\,keV}, from a 20$''$ circular region and applying the same selections to {\it Swift}/XRT observation (00031081002) of SN\,2008D, we obtain a probability of 35\% that the arrival times of the events collected during the
transient \src\ and the X-ray flare associated to SN\,2008D are drawn from the same distribution.
We also note that the luminosity of the X-ray emission detected in the decaying part of the \emph{Swift}/XRT light curve of SN\,2008D \citep{soderberg08} is well below the 3$\sigma$ upper limit we set on the persistent X-ray emission of \src. This means that, if present, a tail similar to that of SN\,2008D would have escaped detection in the \xmm\ data. 

Since we discovered the X-ray transient several years after the event, we had no chance to perform follow-up optical observations to search for possible supernova light. No simultaneous optical data from the Optical Monitor (OM) on board \xmm\ are available, since the position of the transient was outside its field of view, which is smaller than that of the EPIC cameras. However, we found archival optical data for the host galaxy from a monitoring with the Catalina Real-time Transient Survey \citep[more specifically, the Siding Spring Survey;][]{drake09}, including observations performed about two months after the burst.
These data are consistent with a steady source, but they are not sensitive enough to exclude the presence of a supernova, since a supernova as bright as SN\,2008D \citep{soderberg08} at $z=0.092$ would have peaked at $V\approx20$--21\,mag, which could have been detected only in much deeper exposures.

We searched for possible counterparts in other surveys of SNe and transients in the southern sky. Unfortunately, several major SN surveys started a few months or years after our event (for example, La Silla-QUEST, \citet{Baltay13}; DES,\footnote{See the Dark Energy Survey at \url{http://darkenergysurvey.org}.}; SkyMapper Southern Sky Survey, \citet{wolf18}; ESO VST SUDARE\footnote{See the ESO Messenger article at \url{http://www.eso.org/sci/publications/messenger/archive/no.151-mar13/messenger-no151-29-32.pdf}.}). We also checked that the field of \src\ was not covered by the OGLE\footnote{See \url{http://ogle.astrouw.edu.pl/}.} project \citep{udalski15}.
The Transient Name Server\footnote{See the Transient Name Server at \url{https://wis-tns.weizmann.ac.il}.} (TNS) reports four transients in a $2^\circ$ neighbourhood of \src, all of which occurred at least 5 years after our event.
Among the optical SNe in a time window compatible with \src, the closest in sky distance is SN\,2011eb (found in NGC\,782, at $\rm RA=01^h57^m36\fs6$, $\rm Dec=-57^{\circ}48'00\farcs8$ (J2000); \citealt{parrent11}), which is, however, more than $5^\circ$ away from the \xmm\ event, and therefore incompatible.

\subsection{Other interpretations}

The similarities with the X-ray transient associated to SN\,2008D suggest that also in the case of \src\ the X-ray emission from the shock break-out of a core-collapse supernova was detected, but we examined other possible interpretations of the event.  Considering that in the full EXTraS analysis we discovered only a few transients with a duration in the 100--500\,s range, the possibility of chance alignment of a Galactic transient (e.g., an optically-faint flaring star) with a star-forming galaxy at $z\sim0.1$ seems remote. 
Following \citet{bloom02}, we estimate that the probability to find by chance a galaxy as bright as the proposed host or brighter within 2$''$ from the EXTraS transient is $2.5\times10^{-3}$.
Moreover, the evidence for an absorption in excess to the total Galactic $N_{\rm H}$ in that direction derived from the spectral analysis (in particular, for the power-law model) is an additional indication that the X-ray transient is located outside our Galaxy.

Although the X-ray transient position coincides with a peripheral region of the galaxy (possibly the arm of a tidally disturbed spiral galaxy), it is consistent at 1$\sigma$ with the center of the galaxy. However, we can rule out the possibility of a flare from an active galactic nucleus (AGN) since there is no evidence for AGN activity either in optical or at X-rays 
($L_{\rm X}<2.8\times10^{41}$\,\lum). A 5-minutes flare from a quiescent supermassive black hole cannot be excluded, but it would be an unprecedented phenomenon, since the maximum energy released in this kind of X-ray flares by Sagittarius\,A$^\star$ is several orders of magnitude smaller \citep[e.g.][]{ponti17}.

The luminosity of the event is consistent with a tidal disruption event (TDE; e.g. \citealt{burrows11,komossa15}). However, the fast rise of the emission would require a rather exotic scenario: a white dwarf tidally disrupted by an intermediate-mass black hole ($<$$10^5$\,$M_\odot$; eq.\,23 of \citealt{stone16}, see also \citealt{jonker13,glennie15,bauer17}). Another possibility is that the flare is associated to a shock break-out of a star in the course of a `standard' tidal disruption event, before the onset of the accretion (e.g. \citealt{guillochon09}). However, both possibilities are disfavoured against a supernova shock break-out by the rate of tidal disruption events, which is much lower than that of core-collapse supernovae (see Fig.\,\ref{rates}).

Another X-ray bright event with a timescale of $\sim$5 minutes is the pulsating tail of a giant flare from a magnetar \citep[e.g.][]{kaspi17,esposito18}. However, we do not detect any bright initial spike 
and the tail energy is typically $\sim$10$^{44}$ erg, about two orders of magnitude smaller than that emitted by \src.

\subsection{Event rate}
To determine the event rate of X-ray flares as the one detected in \src, we have to evaluate the sensitivity of our search to this kind of events at different distances. Many instrumental effects and observation properties can strongly affect the sensitivity of our search: 
the intensity of the time-variable particle background, chip gaps and defects, instrument settings (operating mode and filter), the presence of bright and extended sources, Galactic interstellar absorption, and the transient spectrum, light curve and off-axis angle. We therefore decided to evaluate the detection efficiency of our search algorithm by simulating the X-ray flare at different flux levels and detector positions and, after the addition of the simulated events to real EPIC data, applying the EXTraS pipeline to see how many of them are recovered as function of the distance.

Since the count statistics of the X-ray flare of \src\ is rather poor, but the transient seems to us to be an analogous of the SN\,2008D X-ray flare, which is much better characterized \citep{soderberg08}, we adopted the spectrum and the light-curve shape of the latter as a template for simulating the flares. In the spectral model of each simulated X-ray flare, we assumed the Galactic absorption expected at its sky coordinates from the survey by \citet{kalberla05}. The positions of the simulated transients were randomly distributed in a square region with an area of 0.324 square degrees, containing the full EPIC field of view. All the relevant instrumental properties (including the point-spread function, as well as vignetting, filter transmission, and detector efficiency effects) were taken into account.

We simulated $\sim$100,000
 transients with 0.3--10\,keV fluence ranging from 10$^{-10}$ to 5$\times10^{-9}$\,erg cm$^{-2}$ and added the simulated photons to the pn and MOS data of $\sim$2,900 randomly-selected \xmm\ observations, corresponding to $\sim$40\% of the 3XMM-DR5 observations.  We then used the EXTraS transient pipeline (Sect.\,\ref{procedure}) to detect the simulated flares, adopting the same detection threshold ($\rm DET\_ML > 15$).
 
\begin{figure}
\centering
\resizebox{\hsize}{!}{\includegraphics[angle=0]{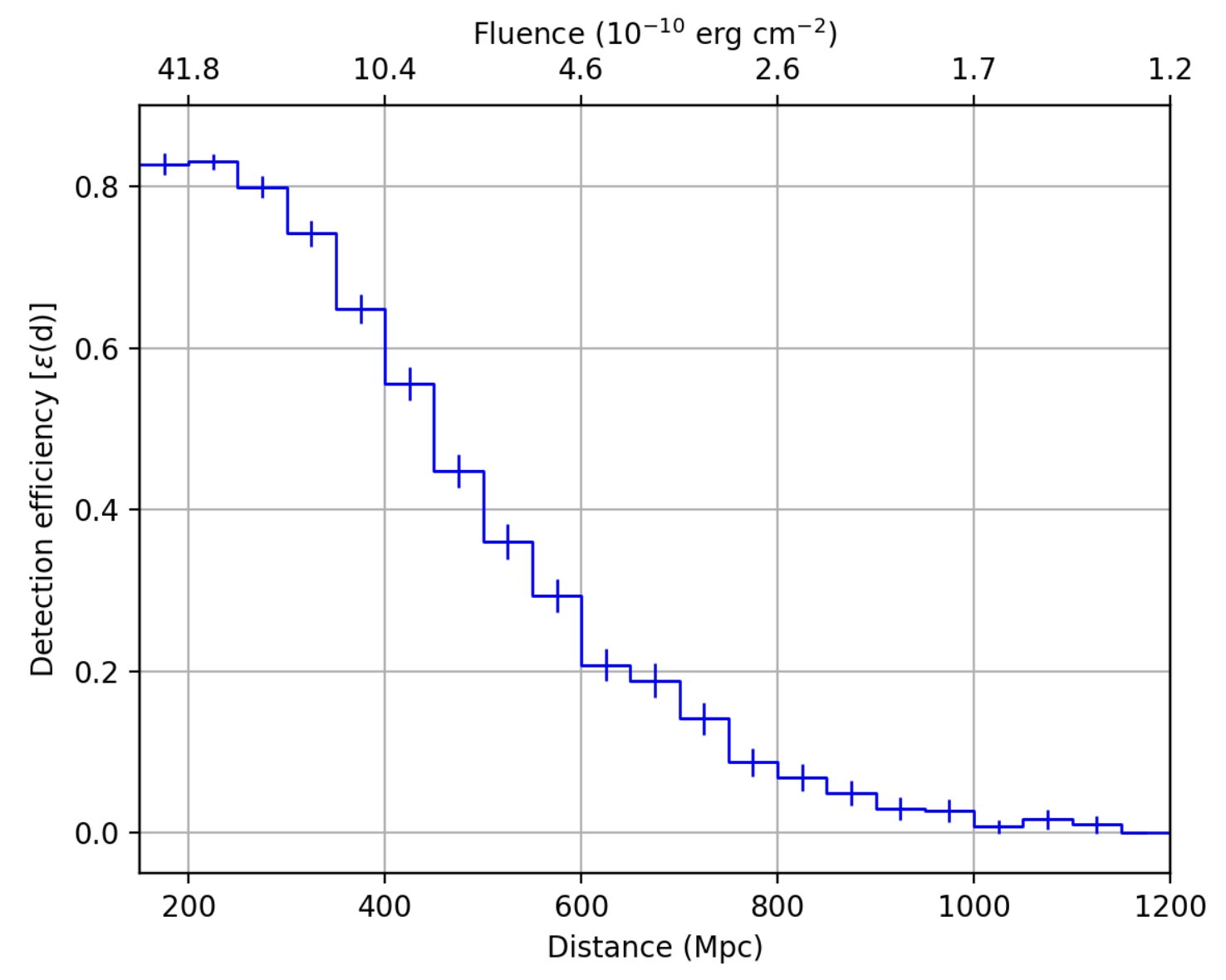}}
\caption{\label{figEFF} Detection efficiency as a function of the simulated source distances. The upper X axis shows the corresponding 0.3--10\,keV fluence of the simulated sources.}
\end{figure}

After filtering out the events simulated outside the field of view and in time intervals during which the instruments were not operating, 48,166 simulated sources had at least 1 valid count. 
We detected significantly 71\% of them. In Fig.\,\ref{figEFF}, we show the success rate ($\epsilon(d)$) as a function of the luminosity distance, derived from the fluence of the simulated sources assuming a flare energy of 2$\times$10$^{46}$\,erg \citep{soderberg08}.

 This detection efficiency $\epsilon(d)$ can be used to compute the effective volume ($V_{\rm eff}$) covered by our survey as:
\begin{equation}
V_{\rm eff} = \sum \epsilon(d_i) V(d_i)= (4.1\pm0.1)\times10^8\,{\rm Mpc}^{3}
\end{equation}
where $V(d_i)$ is the comoving volume in the redshift interval corresponding to the $d_i$ distance bin in Fig.\,\ref{figEFF}. 

The survey coverage of the EXTraS search for transients can be derived from the sum of the exposure maps\footnote{The exposure maps were not corrected for the vignetting effect and in case of simultaneous observations by more EPIC cameras, the one with the largest value was selected.} of all the observations included in the search. The total survey coverage is 1.1 deg$^2$ yr, corresponding to a coverage of the full sky for $T = 2.7\times10^{-5}\,\mathrm{yr} \simeq 14$\,minutes.

\begin{figure*}
\centering
\resizebox{.8\hsize}{!}{\includegraphics[angle=0]{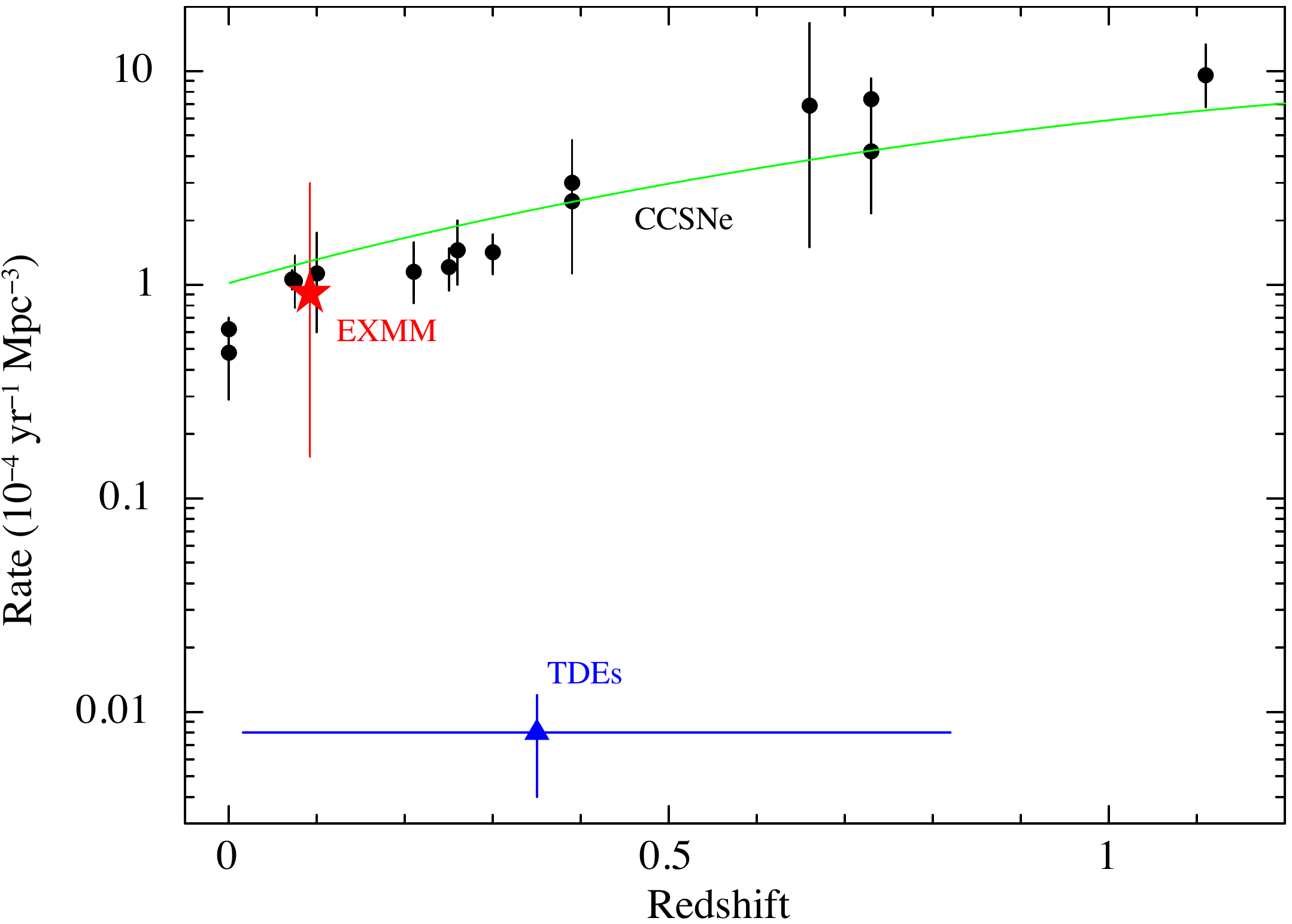}}
\caption{The core-collapse supernova data (black dots) are from \citet[][and the references therein]{cappellaro15}. The green solid line shows the predicted core-collapse supernova rate from \citet{madau14} for a Salpeter initial mass function with the lower and upper mass limits for the progenitors of 8 and 40\,$M_\odot$. The blue triangle represents the rate of tidal disruption event (from \citealt[][]{vanvelzen18}).}
\label{rates}
\end{figure*}

The event rate for $n=1$ detection is therefore:
\begin{equation}
\frac{n}{T\times V_{\rm eff}}=(0.9^{+2.1}_{-0.7})\times10^{-4}\,{\rm 
yr}^{-1}\,{\rm Mpc}^{-3},
\end{equation}
where the 1\,$\sigma$ statistical uncertainty was computed according to \citet{gehrels86}.
This rate is consistent (albeit within large uncertainties) with the core-collapse supernova rate by \citet{cappellaro15} in the $z<0.2$ range sampled by our survey (see Fig.\,\ref{rates}).

It is worth noting that the host galaxy of EXMM\,023135.0--603743 has a small mass and a high specific star formation rate, as expected for the majority of galaxies hosting core-collapse supernovae \citep{botticella17}. Also the distance of $\sim$400\,Mpc is consistent with expectations, as a combination of volume increase and detection efficiency decrease (Fig.\,\ref{figEFF}) at larger distances.
\citet{soderberg08} observed that the X-ray detection of SN\,2008D was compatible with the possibility that  core-collapse supernovae emit this kind of X-ray flares.
Subsequent observations and studies \citep[e.g.][]{mazzali08,modjaz09}, associated the X-ray flare of SN\,2008D to an early cocoon from a massive helium star. In any case, regardless of the exact nature of the transient X-ray emission, the serendipitous discovery of \src\ in a field galaxy rather than in the target of an observation, as it was the case of SN\,2008D in the supernova-rich galaxy NGC\,2770, allows us to derive a more straightforward and unbiased estimate of the rate of such events.

Transients alike \src\ will become detectable up to significantly larger distances with the \emph{Athena X-ray Observatory} \citep{barret19}. Thanks to its $\sim$10 times larger effective area with respect to \xmm, we expect \emph{Athena} will push the 50\% detection efficiency for this kind of events from $z\sim0.1$ to $z\sim0.28$, increasing by a factor $\sim$20 the accessible volume. In particular, the Wide Field Instrument (WFI) will have a field of view $\sim$2.5 times larger than EPIC and, therefore, \emph{Athena} will be able to detect similar X-ray flares $\sim$50 times more frequently than \xmm, which corresponds to more than 2 events per year considering equal observing time shares between the WFI and the X-IFU instruments. 

A large number of events in the local Universe could be detected by soft X-ray detectors with very large fields of view. For example, the \emph{THESEUS} mission \citep{amati18} should be able to detect $\sim$4 supernova shock break-outs per year within 50\,Mpc in its 1\,sr field of view. The accumulation of a significant number of events at different distances will soon allow us to measure the supernova rate in the X-ray band and its evolution with redshift up to $z\sim0.3$ with a precision comparable to present measurements in optical and infrared, and with the advantage of a much smaller bias against supernovae in dusty environment.

\acknowledgments
This research has made use of data produced by the EXTraS project, funded by the European Union's Seventh Framework Programme under grant agreement n.\,607452.
The scientific results reported in this article are based on observations obtained with \xmm, an ESA science mission with instruments and contributions directly funded by ESA Member States and NASA.
This work also used observations at Cerro Tololo Inter-American Observatory (CTIO), National Optical Astronomy Observatory, 
which is operated by the Association of Universities for Research in Astronomy (AURA) under a cooperative agreement with the National Science Foundation. We thank Steven Heathcote for granting us CTIO director's discretionary time and the CTIO staff for performing the observations.  
Part of the funding for GROND (both hardware as well as personnel) was generously granted from the Leibniz-Prize to Prof. G. Hasinger (DFG grant \mbox{HA 1850/28-1}). This work is partly based on tools produced by \mbox{GAZPAR} operated by CeSAM-LAM and IAP.
GN, PE, AT, RS, ADL, PDA, MS, EP, SC, and SM acknowledge funding in the framework of the ASI--INAF contract \mbox{n.\,2017-14-H.0}.
We acknowledge the INAF computing centres of Osservatorio Astrofisico di Catania and Osservatorio Astronomico di Trieste for the availability of computing resources and support under the coordination of the CHIPP project. 
We are grateful to Franz E. Bauer, Ofer Yaron, and Andrzej Udalski for useful information, and to Peter Jonker and Maryam Modjaz for comments on the manuscript.

\facilities{\xmm\ (EPIC), CTIO:\,Blanco\,4-m, MPG/ESO:\,2.2-m/GROND}
\software{SAS \citep{gabriel04}, FTOOLS \citep{blackburn95}, XSPEC \citep{arnaud96}, GOSSIP \citep{franzetti08}, IRAF \citep{tody93} Le\,PHARE \citep{arnouts99}}
 
\bibliographystyle{aasjournal}
\bibliography{biblio}

\end{document}